\documentclass{PoS}

\newcommand{\be}{\begin{equation}}
\newcommand{\ee}{\end{equation}}

\newcommand{\no}{\noindent}
\newcommand{\ba}{\begin{eqnarray}}
\newcommand{\ea}{\end{eqnarray}}
\newcommand{\bma}{\begin{pmatrix}}
\newcommand{\ema}{\end{pmatrix}}

\newcommand*{\bea}{\begin{eqnarray}}
\newcommand*{\eea}{\end{eqnarray}}

\title{Exploring Higgs Sector Spectroscopy}

\ShortTitle{Higgs Sector Spectroscopy}

\author{\speaker{Axel Maas}\thanks{Supported by the DFG under grant number MA 3935/5-1}\\
        E-mail: \email{axelmaas@web.de}}
\author{{Tajdar Mufti}\thanks{Supported by the DFG graduate school 1523-1 and under grant number MA 3935/5-1}\\
        E-mail: \email{tajdar.mufti@uni-jena.de}\\
\\
        Institute of Theoretical Physics, Friedrich-Schiller-University Jena, Max-Wien-Platz 1, D-07743 Jena, Germany} 

\abstract{The Higgs sector of the standard model is field-theoretically a very interesting theory. Because strong and weak coupling domains are continuously connected, only quantitative changes distinguish the various regions. Especially, this is true for the asymptotic spectrum, which can only consist out of gauge-invariant composite, i.\ e.\ bound, states. Since in some regions of parameter space even Regge trajectories are expected to exist, there is immediately the possibility that resonances may also be present in the parameter region characteristic of the standard model Higgs sector. This possibility is discussed in some detail, starting from the definition of the theory to spectroscopy, including excited state analysis, to some considerations whether this could have experimental consequences. The strongest limitation for this exploration turns out to be that the gauge coupling without fermions runs much faster, and the gauge sector is therefore potentially affected.}

\FullConference{31st International Symposium on Lattice Field Theory - LATTICE 2013\\
		July 29 - August 3, 2013\\
		Mainz, Germany}

\begin{document}

\section{Introduction}

The Higgs sector of the standard model contains three ingredients. These are two flavors of Higgs fields, gauged with a non-Abelian SU(2) gauge symmetry with (without QED degenerate) $W$ and $Z$ gauge bosons, and a Higgs self-interaction potential. This theory has a number of remarkable features, if it exists at all. However, the possible problem of triviality will in the following be considered only as the necessity of introducing a (lattice) cutoff, under the assumption that any consequence will be sub-leading for the low-energy regime.

This theory has then the remarkable feature of a quantum phase diagram which is everywhere continuously connected \cite{Fradkin:1978dv}, irrespective of the coupling strengths. Since the local gauge symmetry must be unbroken throughout \cite{Elitzur:1975im}, this implies that the asymptotic spectrum of the theory must be made up of gauge-invariant, composite states \cite{Frohlich:1981yi}, i.\ e.\ bound states. Since for some values of the (bare lattice) couplings an intermediate linear rising potential is observed \cite{Knechtli:1998gf}, even Regge trajectories should be expected. Hence, the possibility of resonances exists.

The interesting question is therefore, whether such resonances could possibly also exist in the regime where the theory resembles the Higgs sector of the standard model. At very weak bare couplings this does not appear to be the case \cite{Wurtz:2013ova}. Here, the situation at somewhat stronger gauge couplings will be discussed, based on some considerations concerning how the theory can be related to the standard-model physics below.

To eventually make statements about the experimental observability of any bound states, if at all, requires a sequence of steps, starting from the determination of the spectrum to the identification of resonances to finally translate lattice results to signatures in experiments. This sequence is discussed in detail in the remainder of this contribution.

\section{Fixing the theory}

Since the Higgs sector alone is only a small part of the standard model, it is not an entirely trivial question of how to match it to it. As will become obvious, this is a much more complicated question than in QCD, where chiral symmetry breaking damps out essentially most quark dynamics in the infrared.

The standard paradigm to match a lattice calculation to experiment is by identifying bound state masses with the masses of experimental observable states. Because the elementary Higgs and $W$/$Z$ fields have the same mass as the lowest gauge-invariant bound states with the same $J^P$ quantum numbers\footnote{Which has lead to a serious discrepancy in the naming scheme between most of the lattice literature and the PDG naming scheme \cite{pdg}. In the former the gauge-invariant bound states are called Higgs and $W$/$Z$, while in the latter the gauge-dependent elementary states carry these names. Herein, the naming scheme of the PDG will be used, and the bound states denoted by their respective quantum numbers $0^+_1$ and $1^-_3$, where the lower index is the flavor representation.} \cite{Frohlich:1981yi,Maas:2012tj}, two masses can be immediately taken from experiment. Choosing the $W$ in the following is, of course, arbitrary.

\begin{figure}
\centering
\includegraphics[width=0.5\linewidth]{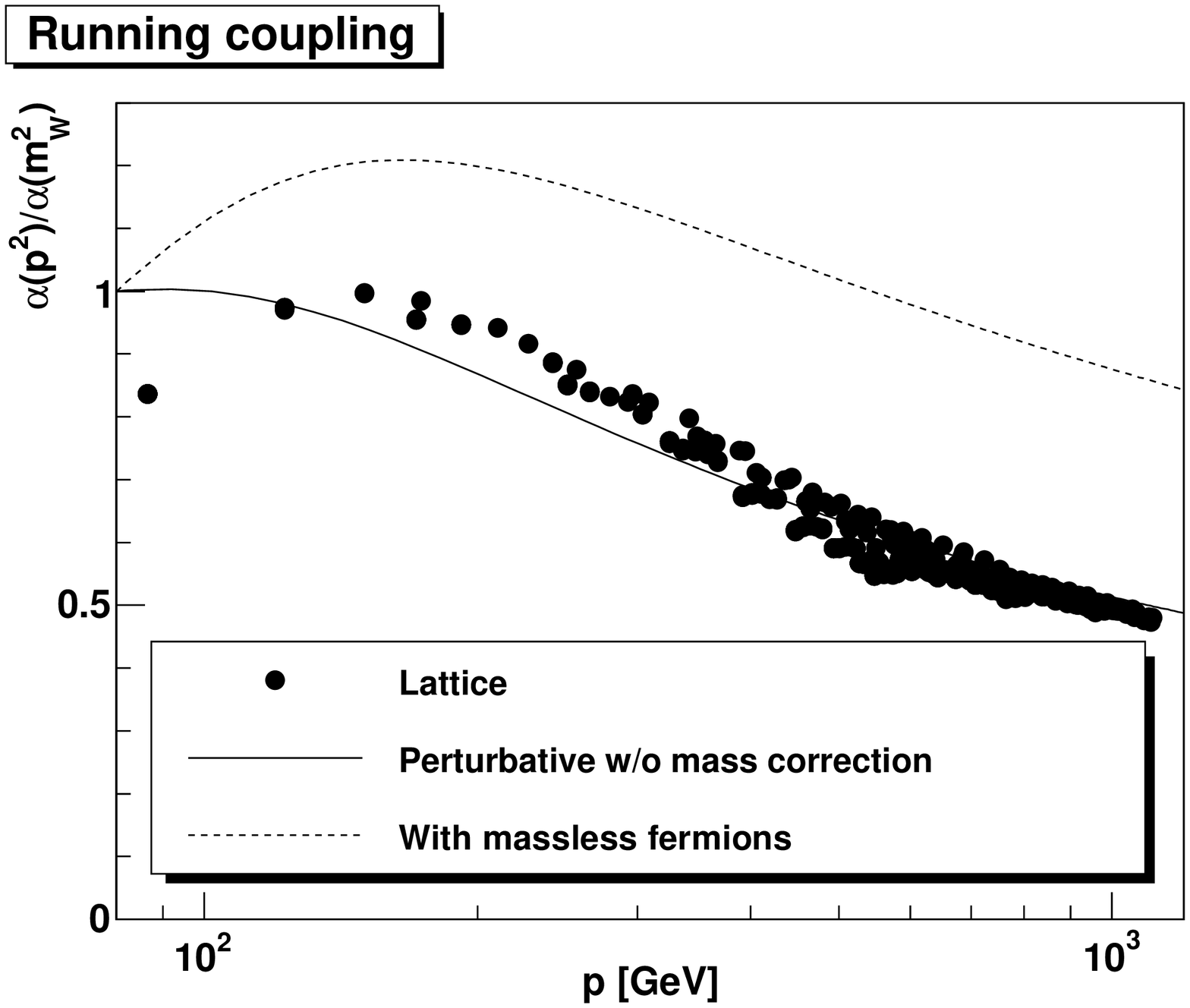}\includegraphics[width=0.5\linewidth]{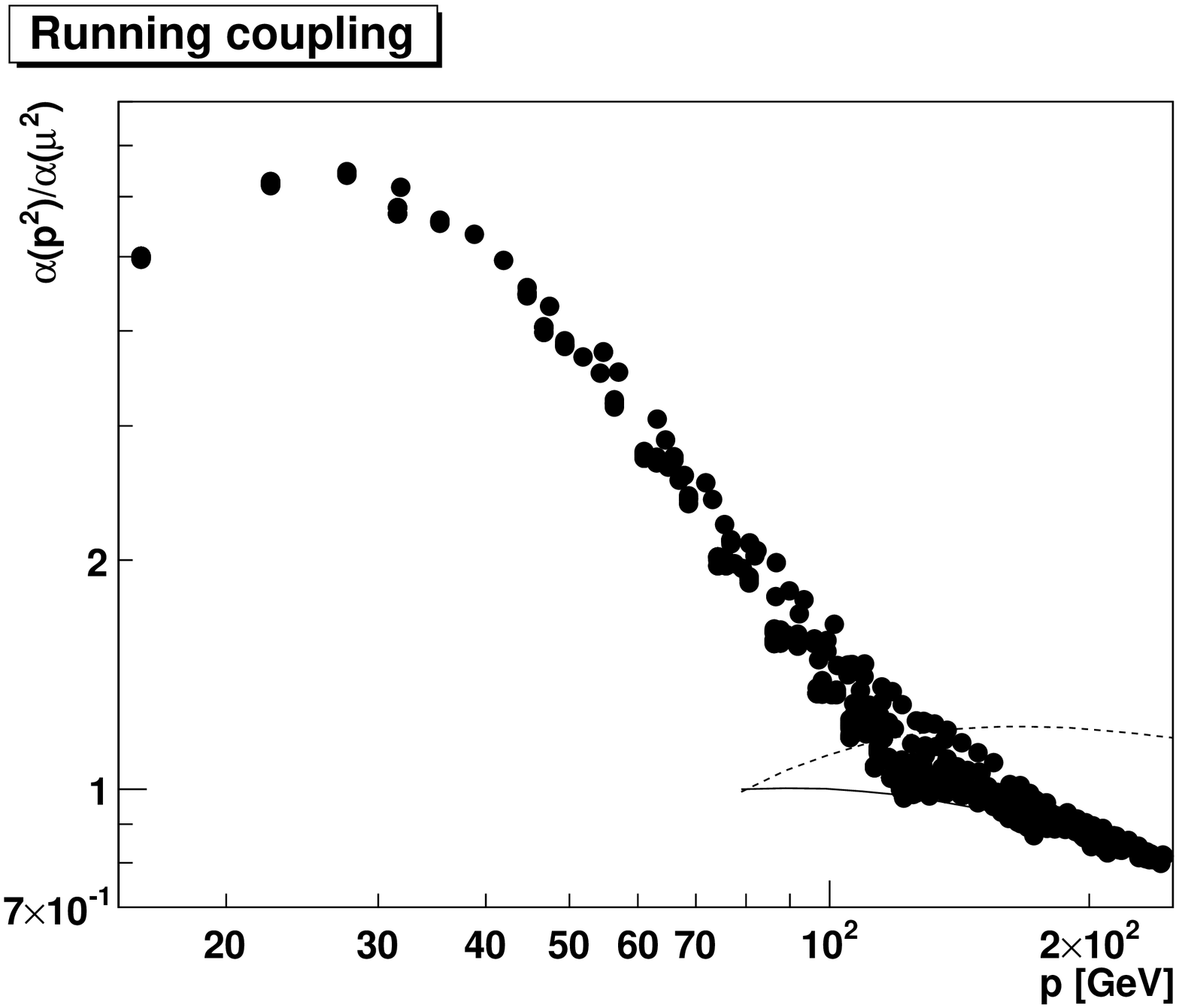}
\caption{\label{alpha}The running coupling in the minMOM scheme inside the would-be Higgs phase (left panel) and inside the would-be confinement phase. The perturbative results only includes leading mass corrections, and fermions are massless. Perturbative results are only shown down to $p^2=(80$ GeV$)^2$. Note that the lattice results are fitted to the high-energy tail, instead of at the $W$ mass, so that the normalization only applies to the perturbative result.}
\end{figure}

Unfortunately, this exhausts the known masses of this theory. Also, for the standard model case of 80 GeV and 125 GeV these states are stable without photons or fermions. Hence, no decay width can be used as an additional input. The only remainder information can thus come from the coupling strengths. The simplest one to access is the running gauge coupling, which can be determined in the miniMOM scheme straightforwardly using two-point functions \cite{vonSmekal:2009ae}. The result is shown and compared to the qualitative asymptotic leading-order perturbative behavior in figure \ref{alpha}\footnote{All results are from $24^4$ lattices. Lattice parameters and further details of the calculations can be found in \cite{Maas:unpublished3} and \cite{Maas:2012tj,Maas:unpublished3}, respectively. The preliminary plots here serve only as illustration.}. It is visible that in the would-be Higgs phase the agreement is good, and remaining discrepancies are likely mainly from further mass corrections \cite{Maas:unpublished3}. The situation is drastically different in the would-be confinement phase.

However, the more important point here is that when including the fermions, the coupling is stronger and runs slower. Hence, the use of only the Higgs and $W$ underestimates the gauge interaction compared to the standard model. This is substantially different from the QCD case, where the difference between quenched and unquenched case is substantially smaller at energies below a few GeV. This implies already that any choice of the third parameter can never make up for the qualitative difference. Quantitatively reliable results require to include fermions.

The aim here is therefore to investigate several different lines-of-constant physics, having the same ratio of Higgs and $W$ mass, to identify generic traits. Eventually, fermions are required for quantitative statements, but this will likely require to use off-lattice methods, given the inherent problems in formulating a parity-violating gauge theory on the lattice.

\section{Spectroscopy}

\begin{figure}
\centering
\includegraphics[width=0.5\linewidth]{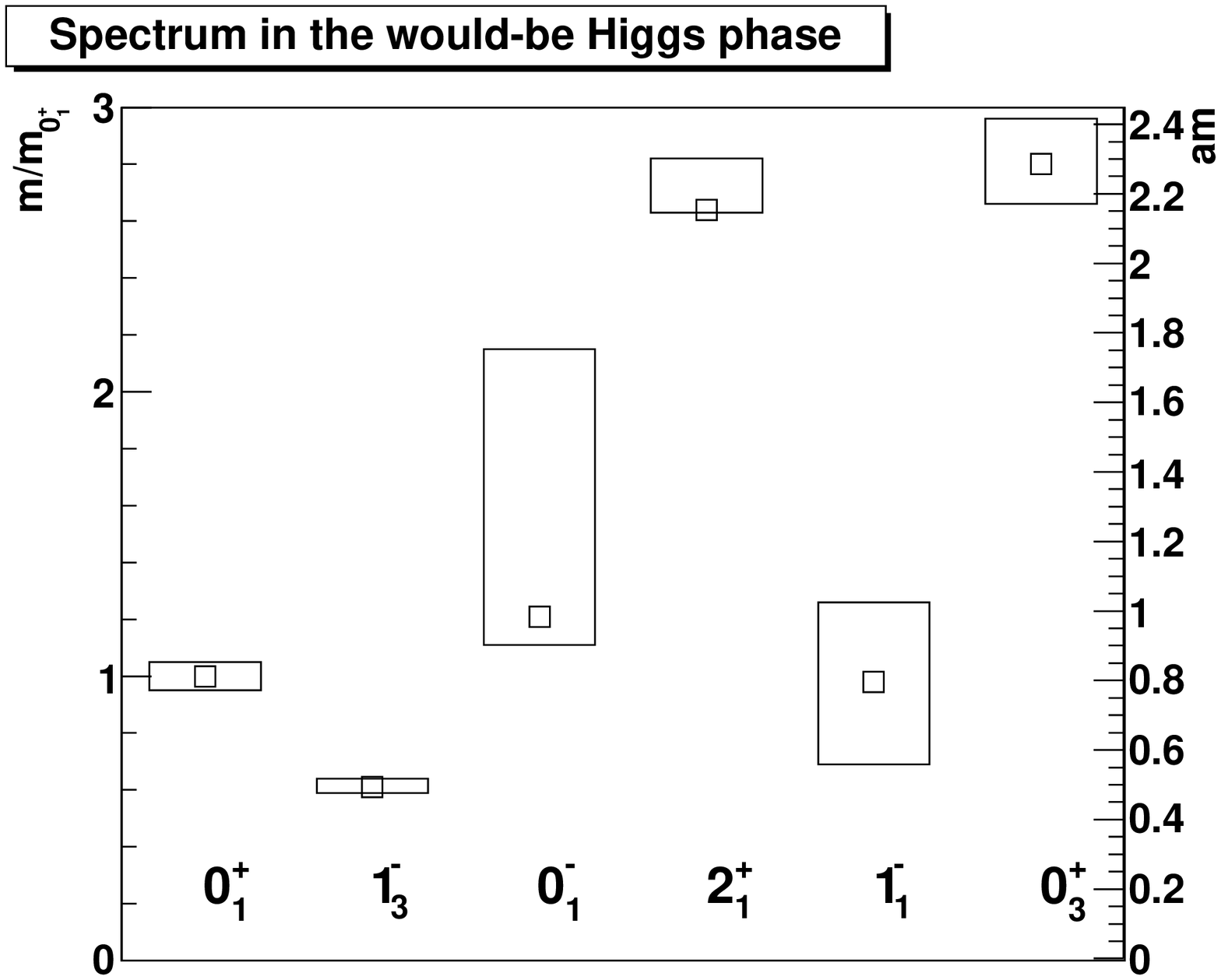}\includegraphics[width=0.5\linewidth]{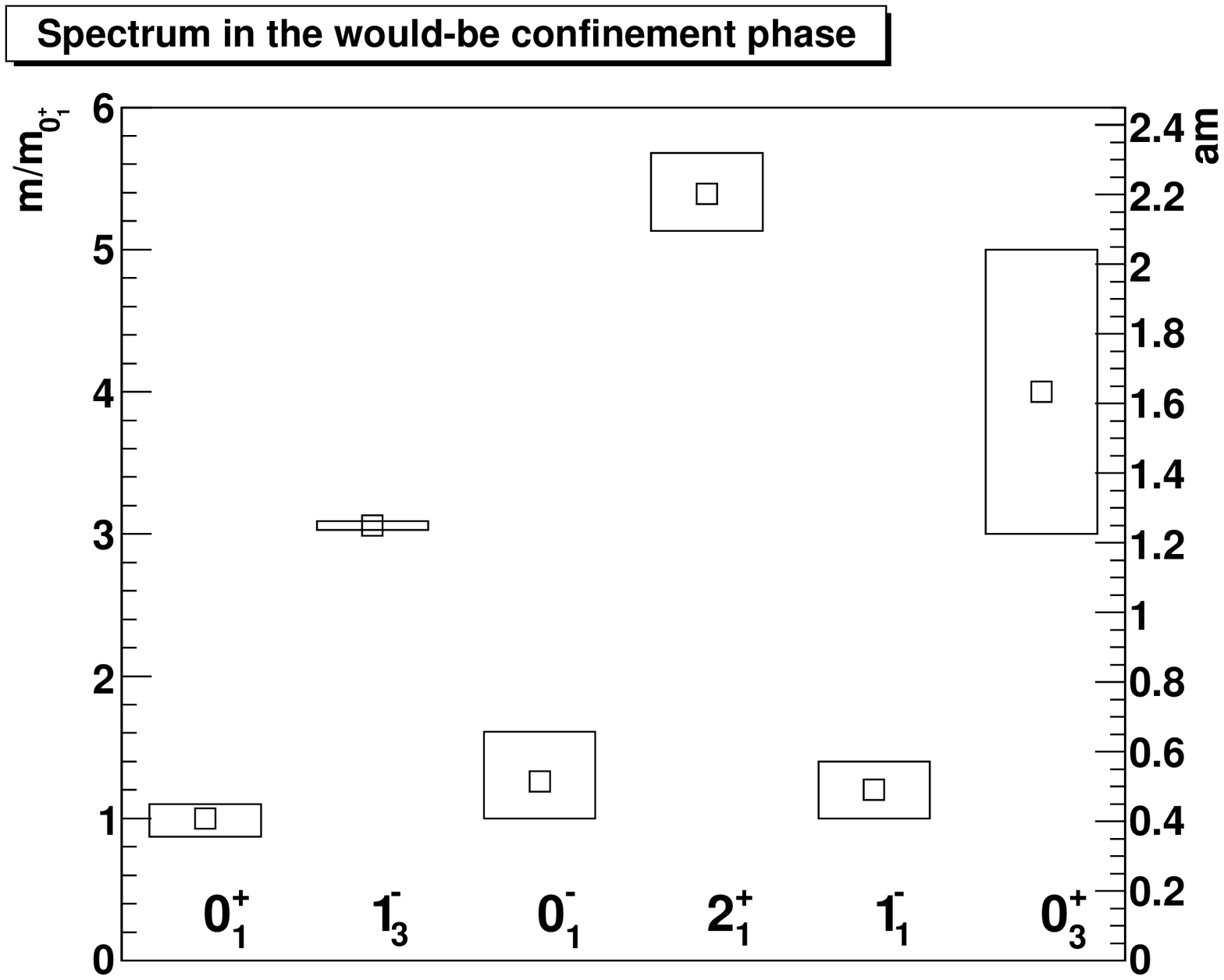}
\caption{\label{spectrum}The spectrum for states with quantum numbers $J^P_f$, where $f$ indicates the flavor representation. Box sizes are statistically errors only, and the small squares are the average values.}
\end{figure}

This still requires to deal with the problem that there is no physical distinction between the would-be Higgs phase and the would-be confinement phase. Figure \ref{spectrum} shows sample results for the spectrum in both phases. In agreement with earlier results \cite{Evertz:1985fc}, the would-be Higgs and would-be confinement phase differ by the ordering of the $1^-_3$ and $0^+_1$ states, the former being lighter in the would-be Higgs phase and heavier in the would-be confinement phase.

\begin{figure}
\centering
\includegraphics[width=0.5\linewidth]{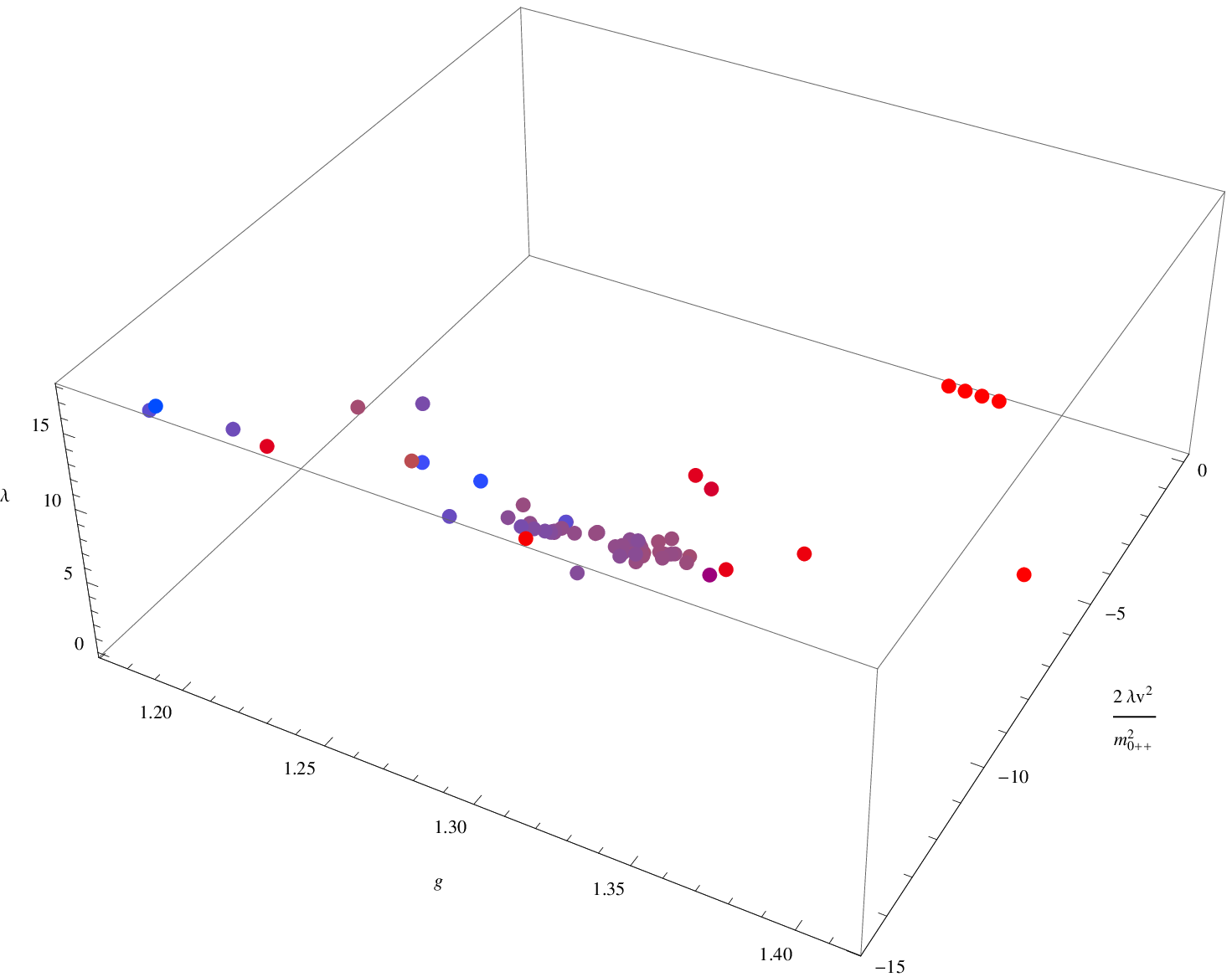}\includegraphics[width=0.5\linewidth]{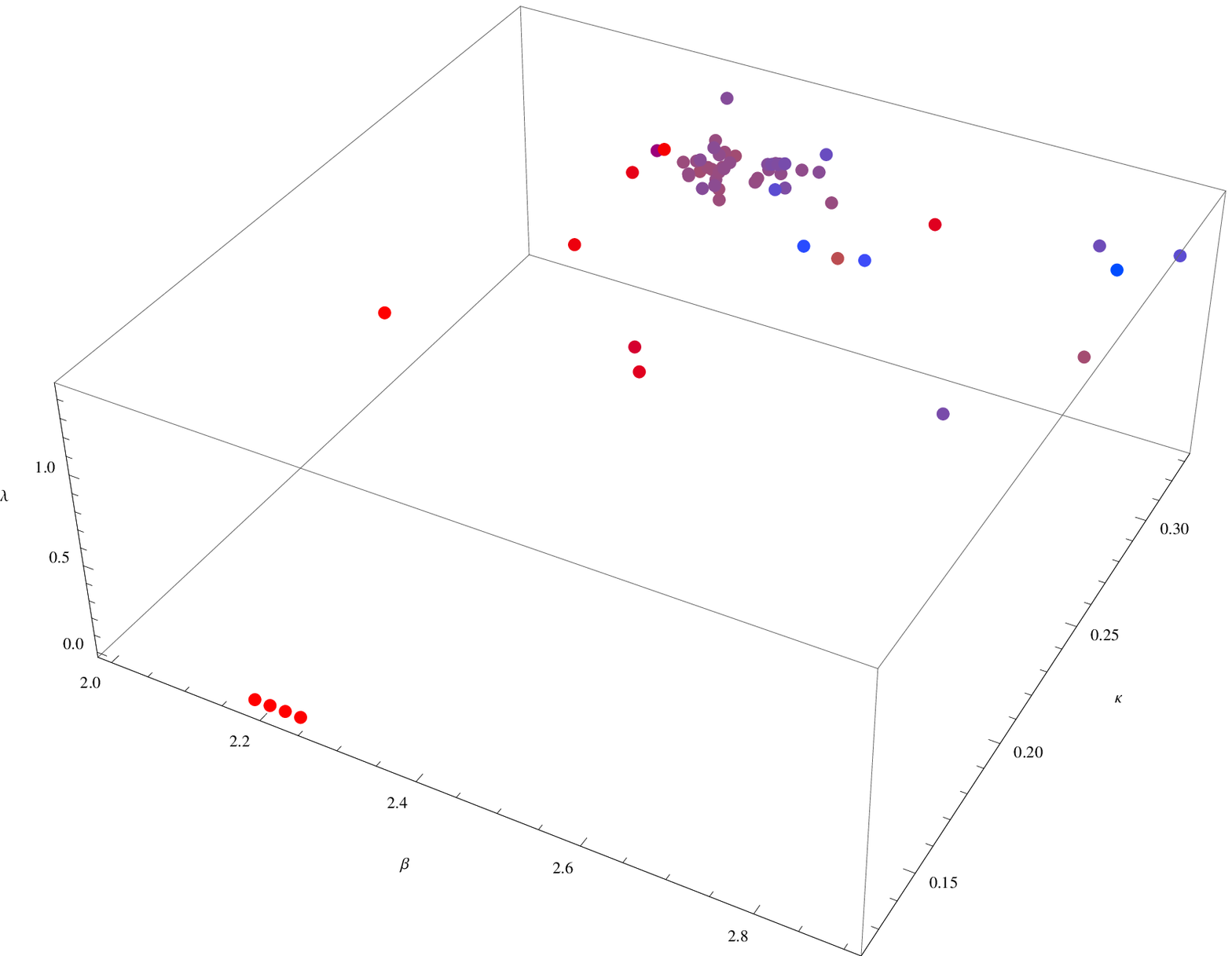}
\caption{\label{lcp}The phase diagram of the Yang-Mills Higgs theory as a function of bare gauge coupling, Higgs 4-point coupling, and the bare Higgs mass in units of the $0^+_1$ mass. Redish points are confinement-like, and blue/purple points are Higgs-like. The darker the points, the smaller is the lattice spacing. The right-hand plot shows the same in terms of the lattice bare parameters of inverse gauge coupling, hopping parameter, and four-Higgs coupling, see \cite{Maas:2012tj} for their definition.}
\end{figure}

The region in the phase diagram where this ordering changes is the same where gauge-dependent observables in Landau gauge \cite{Caudy:2007sf} indicate the change of phases \cite{Maas:unpublished3}, i.\ e.\ this criterion agrees away from the overlap region with the traditional assignment of the Higgs phase by a non-vanishing of the Higgs expectation value in suitable gauges. This criterion will therefore be used here as an operational definition of the phases. The three-dimensional phase diagram using the continuum definition of the coupling constants is shown in figure \ref{lcp}. It is visible that the Higgs phase penetrates as a wedge into the confinement phase, and that also a large negative bare Higgs mass squared can lead to a confinement-like behavior, in contrast to perturbation theory.

\section{Resonances}

\begin{figure}
\centering
\includegraphics[width=0.5\linewidth]{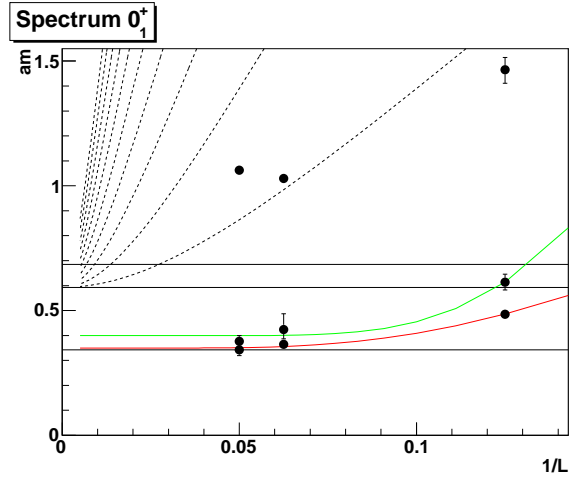}
\caption{\label{lusch}The volume-dependent $0_1^+$ spectrum. The lower horizontal line is the mass for the largest volume, the middle horizontal line is the elastic threshold for the decay into two $1_3^-$ states, and the third horizontal line is the inelastic threshold where the second decay channel into two $0_1^+$ opens up. The dashed lines are the scattering states for two $1_3^-$ particles. The red and green line are exponential fits, see text.}
\end{figure}

In the following, only the Higgs phase will be investigated. Besides the ground states shown in figure \ref{spectrum}, there are also higher-lying states. Their identity is not immediately clear, as whether they can be excited states, resonances, or scattering states. Based on \cite{Luscher:1991cf}, a first hint can come from the volume-dependence of the higher-lying states. For a sample case, this is shown in figure \ref{lusch}. This is obtained from a basis of five operators, including one two-particle operator, and few levels of APE smearing \cite{Maas:2012tj,Maas:unpublished3}. There is only one additional state above the ground state and below the inelastic threshold, though this state is also volume-dependent to a similar extent as the ground state. Further states are far up in the spectrum in the regions where scattering states are expected.

Fitting the visible volume dependence of the ground state with a stable state form \cite{Luscher:1985dn}
\be
m_0(L)=m_0(\infty)+\frac{c_0}{L}\exp\left(-\frac{\sqrt{3}}{2}m_0(\infty)L\right)\label{m0},
\ee
\no yields the red curve visible in figure \ref{lusch}. Similarly, for the next state, which only appears after including scattering states in the operator basis, a similar fit of form \cite{Luscher:1985dn},
\be
m_1(L)=m_1(\infty)+\frac{c_1}{m_1(\infty)L}\exp\left(-\sqrt{m_c^2-\frac{m_1(\infty)^2}{4}}L\right)\label{m1}
\ee
\no is possible, and also shown in figure \ref{lusch} in green. The mass $m_c$ is the one of the $1^-_3$ ground state, assuming these to be the constituents. At the same time, no reasonable fit can be found under the assumption that the state is a scattering state. Of course, larger volumes and finer lattices will be needed to confirm any of these results. Still, taking these results at face value, this seems to indicate the presence of an excited state, which cannot decay in the present theory. This would turn up as a second state in the Higgs channel, with here a roughly  15\% larger mass.

This situation is not generic for the so far investigated cases. E.\ g.\ at slightly different parameters, no such excited state has been observed below the elastic threshold. Whether this is a lattice artifact or due to the employed operator basis, or a genuine effect remains to be seen. However, also in these cases states above the elastic threshold are seen, though for a full phase shift analysis \cite{Luscher:1991cf} to decide whether these are genuine resonances or just scattering states requires significantly more statistics than currently available.

\section{Possible experimental signatures}

Should either such excited states or resonances be confirmed in lattice calculations, it is an interesting question, whether they are experimentally relevant, and, moreover, why the not-observation of any such state has not ruled them out so far. Both of these questions can be answered when contemplating the ground state. Take a suitable gauge, where the Higgs expectation value $v\vec n$ in some direction of charge space $\vec n$ does not vanish. The correlation function for the Higgs-Higgs bound state of the Higgs fields $\phi$ in the $0^+_1$ channel can then be expanded as \cite{Frohlich:1981yi}
\be
\langle\phi_i^+(x)\phi^i(x)\phi_j^+(y)\phi^j(y)\rangle\approx v^4+4v^2(c+\langle\eta^+_i(x) n^i n_j^+\eta_j(y)\rangle)+{\cal O}(\eta^3)\label{correl},
\ee
\no where $\eta$ are the fluctuations of the Higgs fields around its expectation value $\eta=\phi-\vec v$. Thus, to this order in the quantum fluctuations the pole position, and hence the mass, of the bound state and the elementary particle agrees\footnote{Some subtleties are involved when it comes to renormalizing the expressions on the right-hand side.}. This incidentally also explains why the description in terms of the elementary particles as asymptotic states in the Higgs sector works so well, in contrast to QCD.

More important here is that the correlator of the Higgs particle is not aware of excited states, or resonances, and thus has only a single pole, while the bound state correlator on the left hand side does has the corresponding pole structures. They must therefore be hidden on the right-hand side in the sub-leading contributions. Hence, they are suppressed by at least one order of the quantum fluctuations, which for the parameters of the standard model is a small quantity. This already explains why such states have not been observed so far, as they will require quite large statistics.

This leaves the question of how to observe such states. The eigendecomposition of the states indicate that e.\ g.\ the excited state shown above may have a substantial overlap with operators consisting mainly of the gauge bosons. It therefore appears reasonable that such resonances can be formed in the scattering of $W$s or $Z$s, particularly of longitudinal ones. This interferes with quartic gauge interactions. The presence of additional states should hence be visible as an anomalous quartic gauge coupling, an effect also expected in beyond-the-standard-model physics.

To estimate the effect, it is useful to encode the resonances in a low-energy effective theory, e.\ g.\ in terms of gauge-invariant operators of higher dimensions, or non-gauge-invariant operators of lower dimensions, where in either case results will at best be useful at low energies, in the latter case at most at tree-level. Such low-energy effective theories can then be used as input to Monte-Carlo generators.

Results obtained from Whizard \cite{Alboteanu:2008my} have already confirmed that such light additional states, below 300 GeV, have to be coupled very weakly, in agreement with the considerations above, though these results have been obtained in a different context. On the other hand, employing Sherpa \cite{Gleisberg:2003ue}, using only an effective coupling to a two-$W$ state, shows that such a weakly coupled state will not be visible with any reasonable statistics in a 2-to-2 process. This is different in a 1-to-3 process, where there is much less standard-model background, especially as the non-perturbative contribution can provide additional couplings \cite{Maas:unpublished3}. In this case, a sizable signal may exist in certain channels, depending on the precise parameters. This has to be studied further \cite{Maas:unpublished3}.

Of course, it is possible that there is no additional genuine resonance or excited state. Still, some remnant, like with the $\sigma$-meson of QCD, of the states in the continuously connected would-be QCD phase may survive, and also show up in experiments. This would require a very careful phase shift analysis to disentangle this effect from perturbative and other sources.

\section{Conclusions}

The bound state spectrum of Higgs-Yang-Mills theory is a rewarding topic, especially for the case of excited states. In the would-be QCD phase, the conjectured existence of Regge-trajectories is likely an opportunity to study confining properties without the additional complications introduced by chiral symmetry breaking. In the would-be Higgs phase, if remnants of these additional states still exist, these will be additional experimental signatures, and also a possible additional source of background in searches for new physics. Finally, studying different sets of parameters may be useful to understand generic features of more strongly coupled theories of this type, which may turn up as low-energy-effective theories of new physics.

\bibliographystyle{bibstyle}
\bibliography{bib}

\end{document}